\documentclass{article}
\usepackage{spconf,amsmath,epsfig}

\usepackage{url}
\usepackage{graphicx}
\usepackage{epstopdf}

\usepackage{footnote}
\usepackage{balance}
\usepackage{amssymb}

\usepackage{amsmath}
\usepackage{multicol}
\usepackage{multirow}
\usepackage{anyfontsize}
\usepackage{array}
\usepackage{hhline}
\usepackage{float}
\usepackage{enumitem}
\usepackage{tabularx}

\let\OLDthebibliography\thebibliography
\renewcommand\thebibliography[1]{
  \OLDthebibliography{#1}
  \setlength{\parskip}{0pt}
  \setlength{\itemsep}{0pt plus 0.3ex}
}

\begin{document}\sloppy

\def\x{{\mathbf x}}
\def\L{{\cal L}}

\title{The Hyper360 toolset for enriched 360$^\circ$ video  }
%

\name{\begin{tabular}{c}Hannes Fassold$^1$, Antonis Karakottas$^2$, Dorothea Tsatsou $^2$, Dimitrios Zarpalas$^2$, Barnabas Takacs$^3$,\\Christian Fuhrhop$^4$, Angelo Manfredi$^5$, Nicolas Patz$^6$, Simona Tonoli$^7$, Iana  Dulskaia $^8$\end{tabular}}
\address{$^1$ JOANNEUM RESEARCH - DIGITAL\\$^2$ Centre for Research and Technology Hellas (CERTH) - Information Technologies Institute (ITI) \\$^3$  Drukka Kft/PanoCAST\\$^4$ Fraunhofer FOKUS\\$^5$ Engineering Ingegneria Informatica S.p.A.\\$^6$ Rundfunk Berlin-Brandenburg\\$^7$ R.T.I. S.p.A. – Mediaset S.p.A\\$^8$ Eurokleis s.r.l.}

\maketitle

\begin{abstract}
360$^\circ$ video is a novel media format, rapidly becoming adopted in media production
and consumption as part of today’s ongoing virtual reality revolution. Due to its novelty, there is a
lack of tools for producing highly engaging 360$^\circ$ video for consumption on a multitude of platforms. In this work, we describe the work done so far in the Hyper360 project  on tools for 360$^\circ$ video. Furthermore, the first pilots which have been produced with the Hyper360 tools are presented.
\end{abstract}
\section{Introduction}
\label{sec:intro}
Omnidirectional (360$^\circ$) video content recently got very popular in the media industry, because it
allows the viewer to experience the content in an immersive and interactive way. 
A critical factor for the long term success of 360$^\circ$ video content is the availability of convenient tools for producing and editing 360$^\circ$ video content for a multitude of platforms (like mobile devices or VR Headsets). In order to address these shortcomings, in the ongoing Hyper360 \footnote{\url{https://www.hyper360.eu/}} project an innovative toolset for 360$^\circ$ video is proposed.
Specifically, within the Hyper360 project an innovative complete solution for the capturing, production, enhancement, delivery and consumption of an  free viewpoint video to the OTT \footnote{An over-the-top (OTT) media service is a streaming media service offered directly to viewers via the Internet} media sectors is developed and validated through the production of pilots with these tools and the assessment of the  pilots by the audience. By augmenting 360$^\circ$ video with 3D content acting as a \emph
{Mentor}, novel and powerful storytelling opportunities are added in the production process. Additionally, by extracting and adapting to the viewer preferences, a personalized consumption experience can be provided.

The whole Hyper360 pipeline consists of four layers:
\begin{itemize}
  \item  The \textbf{Capturing layer} contains the \emph{OmniCap} tool (described in section \ref{subsec:omnicap}) which is respsonsible for capturing 360$^\circ$ video with a variety of cameras (multi camera arrays, tiny fisheye lense devices, ...) and ensuring high quality content via the integrated quality analysis component.
  \item  The \textbf{Production layer} comprises two  post-processing tools for enriching 360$^\circ$ videos and offering an enhanced viewpoint experience. Specifically, the \emph{OmniConnect} tool (section \ref{subsec:omniconnect}) manages the annotation of 360$^\circ$ media with hyperlinks to other media and content metadata. Additionally, the  \emph{CapTion} tool (section \ref{subsec:caption}) is responsible for capturing the performance of human narrators in 3D and embeddding the generated 3D content in the 360$^\circ$ video.
  \item  The \textbf{Delivery layer} comprises cloud services that facilitate the delivery of  produced, integrated media under seamless, personalised 360$^\circ$ experiences. Within this scope, Hyper360's \emph{OmniCloud} includes  user profiling and recommendation services for personalized consumption (section \ref{subsec:profiling}) and the automatic camera path generator tool for generating a 2D video for playout on a conventional TV set (section \ref{subsec:autocmam}).
  \item  The \textbf{Presentation layer} contains the various players (section \ref{subsec:players}) required for the playback of the enriched 360$^\circ$ content on multiple platforms. 
\end{itemize}
The user partners in the Hyper360 project, the broadcasters Rundfunk Berlin-Brandenburg and Mediaset, have produced pilots (see section \ref{sec:pilots}) using the first prototype of the Hyper360 toolset. Furthermore, they have carried out dedicated audience assessment sessions for the produced pilots.

\section{Tools}
\label{sec:tools}


\subsection{OmniCap}
\label{subsec:omnicap}

\begin{figure}[t]
	\centering
		\includegraphics[width=0.48\textwidth]{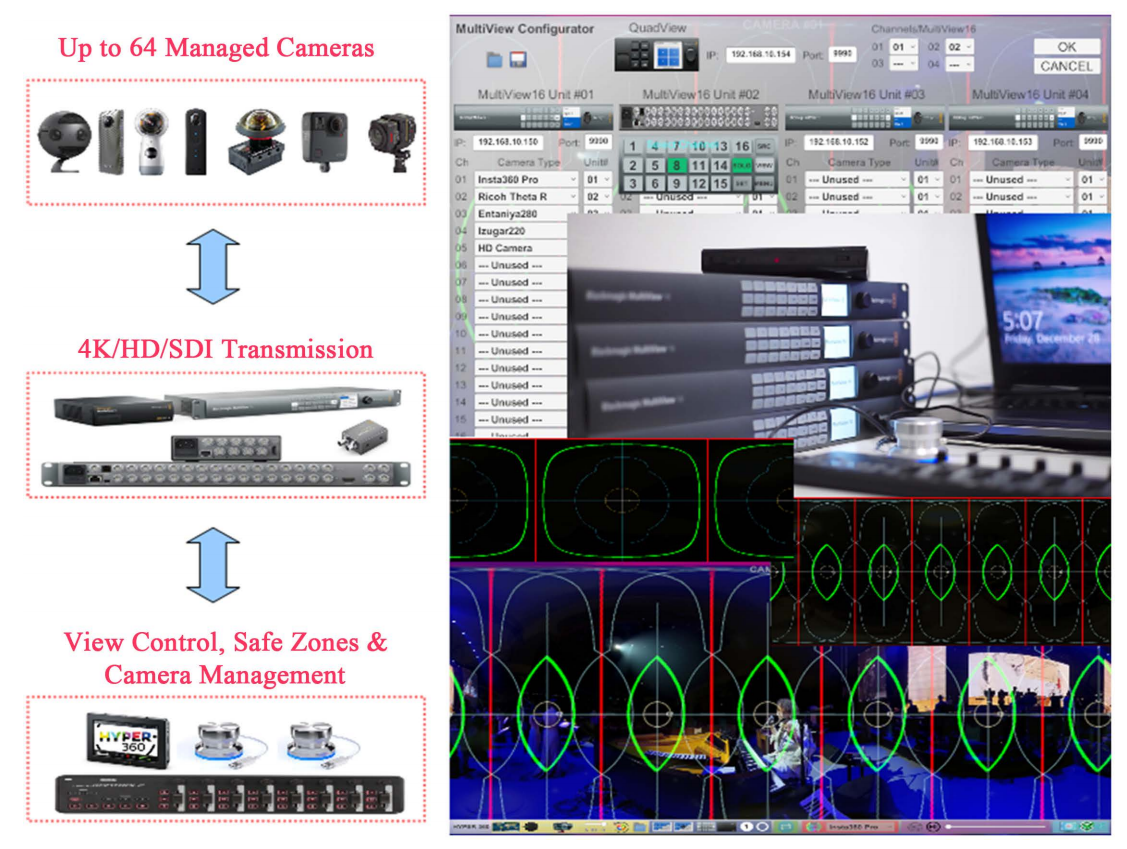}
	\caption{OmniCap tool for a mixed 360$^\circ$ and regular cameras setup.} 
	\label{fig:omnicap}
\end{figure}
The OmniCap tool was designed with the purpose of managing multiple 360$^\circ$ and regular camera types for professional productions. It employs broadcast hardware (4K resolution, 3D stereoscopic capture, Black Magic Design MultiView Units and SDI+Ethernet camera connectivity) and provides a pipeline to manage up to 36 cameras simultaneously placed on a set.  The system supports multiple camera types and configurations in a broad range of mounting and imaging options including shader-based live stitching, safe-zones, camera parameter controls, real-time effects, virtual camera movements and integration of 3D elements into the live scene as required. In Figure \ref{fig:omnicap} its usage for a mixed setup of 360$^\circ$ and regular cameras is shown. In order to support high quality on-set capture as well as post production workflows, the OmniCap system also includes a 360 desktop player with integrated tracking,  object detection and quality control. For the quality control, existing algorithms for detection of defects like signal clipping, blurriness, flicker or noise level on conventional video have been adapted to the specifics of 360$^\circ$ video (especially the equirectangular projection) and extended in order to provide localized defect information. The adaptions for the blurriness algorithm are described in \cite{Fassold2019Blurriness}. General strategies how to adapt computer vision algorithms to the specifics of 360$^\circ$ video (like the equirectangular projection) are given in \cite{Fassold2019Adapting}.

\subsection{OmniConnect}
\label{subsec:omniconnect}
\begin{figure}[t]
	\centering
		\includegraphics[width=0.48\textwidth]{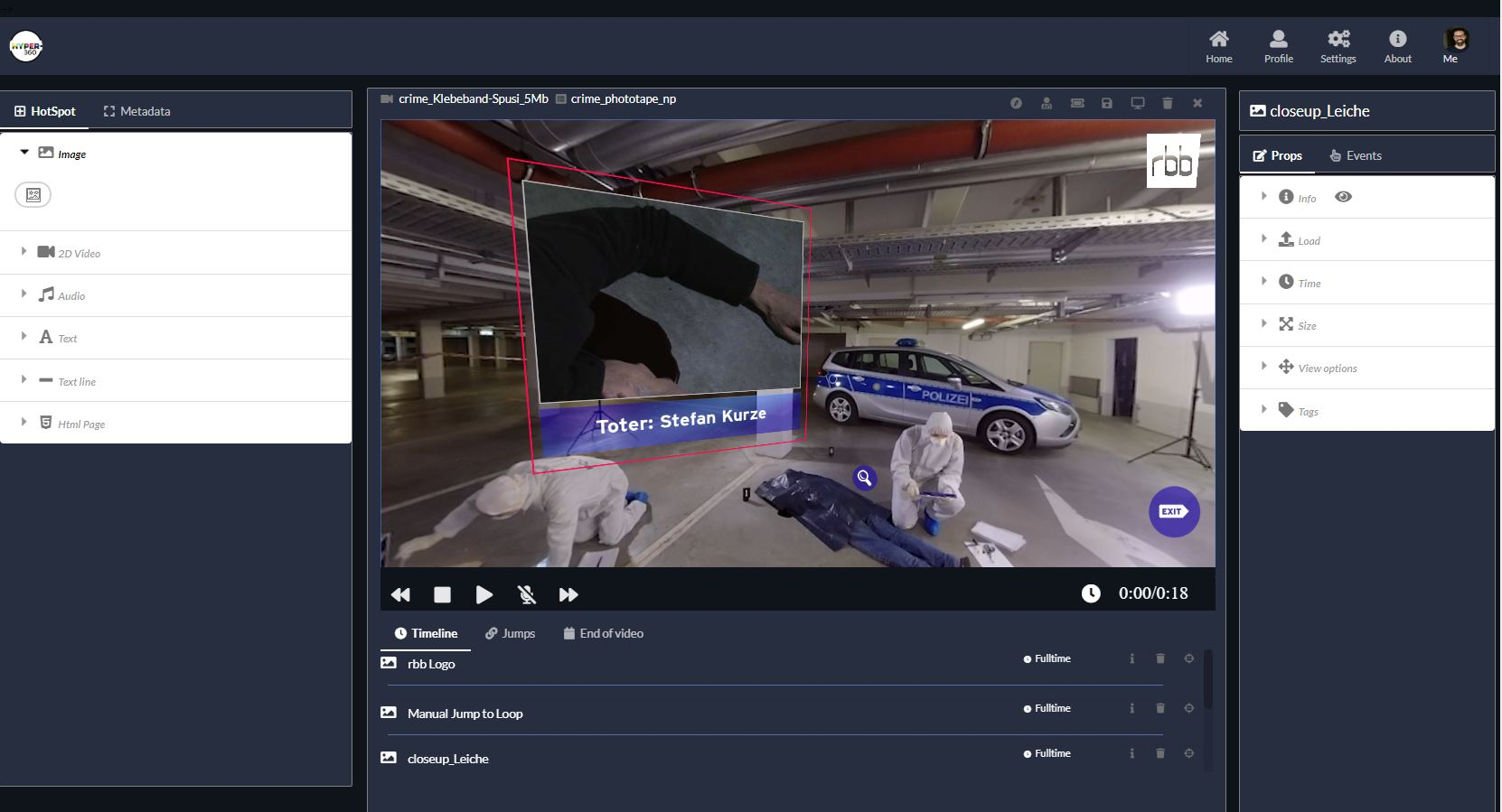}
	\caption{OmniConnect tool for enriching 360$^\circ$ video} 
	\label{fig:omniconnect}
\end{figure}
OmniConnect offers many features to enrich 360$^\circ$ videos with different types of hotspots such as shapes, 2D videos, audio fragments, html pages, multiline texts, images and metadata. To do that OmniConnect employs a user-friendly web interface (see Figure \ref{fig:omniconnect}) compatible with any browser supporting the HTML5 and WebGL standards.  It includes an integrated player for previewing purposes. In addition to that, OmniConnect also offers the possibility to publish the enriched 360$^\circ$ video file by specifying one or more target device classes (iOS, Android, PC and HbbTV).

\begin{figure}[t]
    \centering
    \includegraphics[width = 0.48\textwidth]{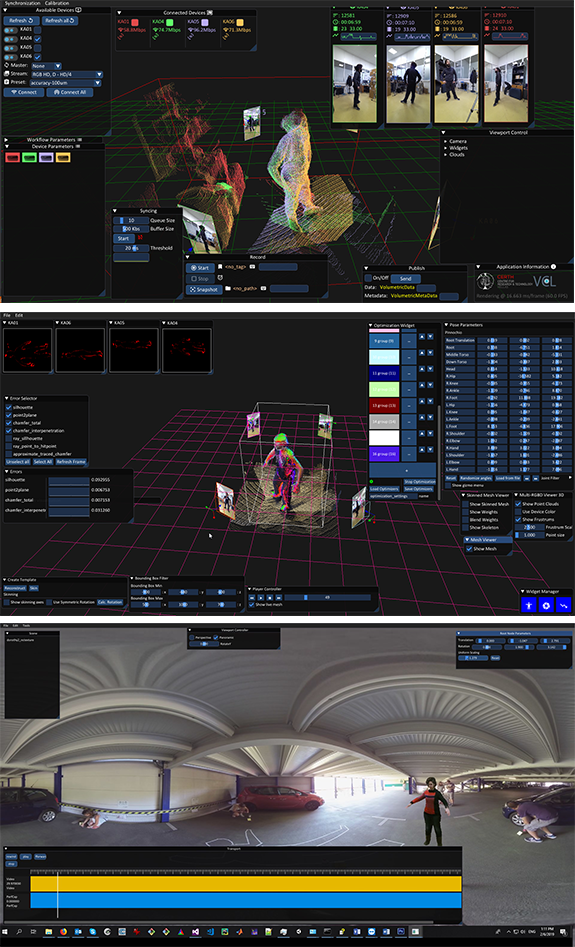}
    \caption{Caption software toolset. From top to bottom: The Volumetric Capture, the Performance Capture and the 360 Fusion applications.}
    \label{fig:CapTion}
\end{figure}

\subsection{CapTion \& 3D Mentor}
\label{subsec:caption}

The CapTion toolkit comprises a set of tools designed for mixed reality media productions that aim to combine two types of Free-Viewpoint Video (FVV), namely 360$^\circ$ and 3D content. The first system of the tool-chain is an affordable and portable volumetric capturing hardware system with its corresponding software application \cite{sterzentsenko2018low}, which gives the ability to record spatially calibrated \cite{papachristou2018markerless} and temporally aligned RGB-D videos. It is followed by the Performance Capture application which produces dynamic (i.e. animated) 3D textured meshes of human performances \cite{alexiadis2018fast}. The exported 3D content can be embedded in omnidirectional media through the 360$^\circ$ Fusion tool that utilizes neural networks for 360$^\circ$ scene understanding \cite{Zioulis_2018_ECCV, karakottas2018360d, zioulis2019spherical, karakottas2019360} to allow for seamless 
placement and realistic rendering of the 3D content inside omnidirectional media.
Figure \ref{fig:CapTion} presents sample views of the Volumetric Capture, Performance Capture and 360$^\circ$ Fusion applications.

\subsection{Profiling and Recommendation Engine}
\label{subsec:profiling}
\begin{figure}[t]
    \centering
    \includegraphics[width = 0.48\textwidth]{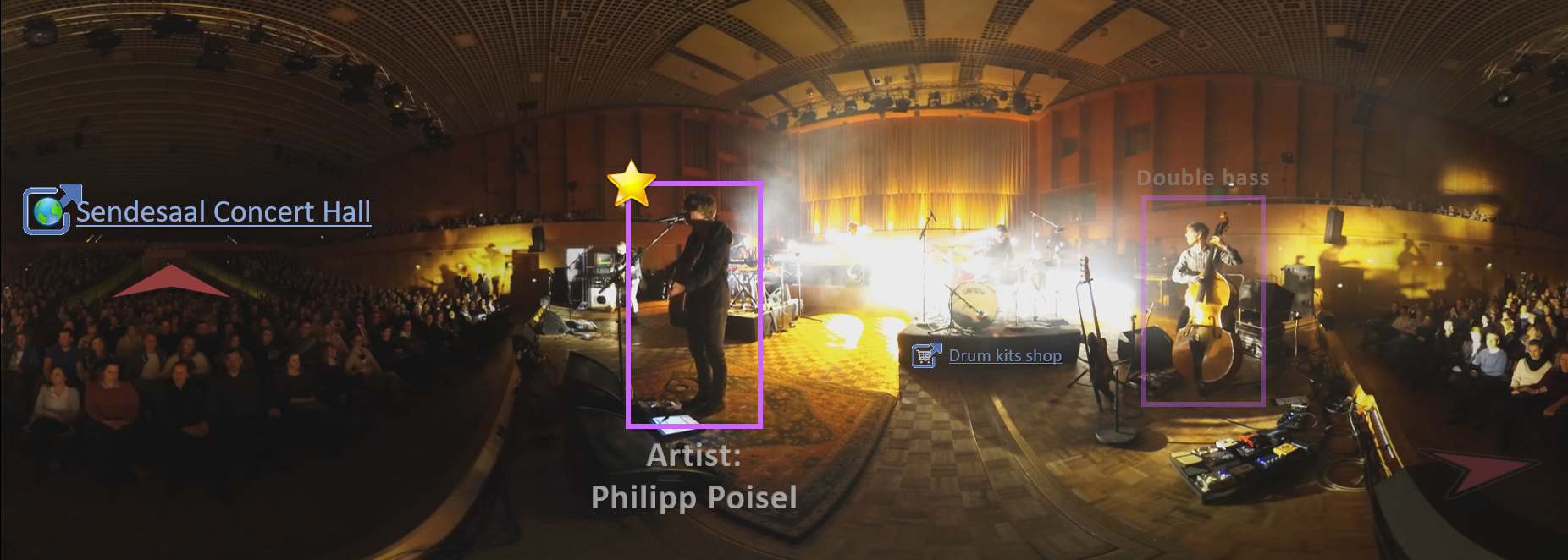}
    \caption{Example of embedded objects recommendation in omnidirectional video.}
    \label{fig:Recom}
\end{figure}

The personalisation mechanism in Hyper360 relies on the holistic representation of content annotation, in order to minimise loss of information and characterise it with enhanced metadata. To this end, through Hyper360's \emph{Semantic Interpreter}, content annotation is construed to a set of fuzzy concept instances, based on an expanded version of the LUMO ontology \cite{tsatsou2014lumo}. This interpretation builds upon learned lexico-syntactic relations between free-form, domain-specific metadata. Subsequently, Hyper360's \emph{Profiling Engine} harvests the novel opportunities that omnidirectional video offers for implicitly capturing the viewers’ preferences. The viewpoint choice allows the \emph{Profiling Engine} to capture spatio-temporal and other behavioural information (which part of the video the viewer focuses on and for how long, which objects in a scene they interact with, etc.), which are combined with appropriate semantic meaning of the content. Lastly, the \emph{Recommendation Engine} semantically matches learned user profiles with semantic content metadata, through an extension of the LiFR fuzzy reasoner \cite{tsatsou2014lifr}. The application of recommendation is three-fold within Hyper360, offering personalised navigation within the 360$^\circ$ media, which is manifested both as cues for personalised camera path(s), as well as cues for personalised 3D Mentor narratives, while also achieving targeted embedded objects (hotspots, hyperlinks, nested media) delivery, as seen in Figure \ref{fig:Recom}.

\subsection{Automatic Camera Path Generator}
\label{subsec:autocmam}
The goal of automatic camera path generation (automatic cinematography / virtual director) is to calculate automatically a visually interesting camera path from a 360$^\circ$ video, in order to provide a traditional TV-like consumption experience. This is necessary for consumption of a 360$^\circ$ video on older TV sets, which do not provide any kind of interactive players for 360$^\circ$ video. Furthermore, even on devices capable of consuming 360$^\circ$ videos interactively, an user might prefer a lean-back mode, without the need to navigate around actively to explore the content. The initial prototype of the automatic camera path generator (more details can be found in \cite{Fassold2019Autocam}) is based on the information about the scene objects (persons, animals, ...), which is extracted with the method given in \cite{Fassold2019Omnitrack}.
For each scene object, a saliency score is calculated based on several influencing factors (object class and size, motion magnitude, neighbours of object), which indicates the “interestingness” of the object. From the calculated saliency scores for the scene objects, an automatic camera path is generated for the current shot by tracking the object with the highest saliency score.

\subsection{OmniPlay}
\label{subsec:players}
The Hyper360 player technology is a suite of players for multiple platforms delivering the backbone of the Hyper360 playback environment by supporting all standard player functionalities, hotspots (images, text, video) as well as a number of innovative functionalities based on the free viewpoint media’s characteristics (like fusion of  360$^\circ$ video and 3D content, automatic camera path, personalisation  multi-viewport coalescing and nested media graphs). Players have been developed for the following platforms: Windows, web browser, smartphones (Android / iOS), VR Headsets like HTC Vive or Oculus Go, SmartTV (HbbTV / AppleTV) and traditional TV. For the implementation of the player, proper software frameworks have been used for the specific platform (mainly Unity, JavaScript / WebGL and XCode). As not every platform supports the same user interaction modality (e.g. SmartTV supports only key-based navigation), platform-specific adaptions to the UI have been done.

\section{Pilots}
\label{sec:pilots}

The user partners in the Hyper360 project, the broadcasters Rundfunk Berlin-Brandenburg (RBB) and Mediaset (RTI), employed the first prototypes of the developed tools in order to produce pilot content. For this, RBB focused on the \emph{Immersive Journalism} scenario, which allows first person experience of the events or situations described in news reports and documentary film. 
The focus of RTI was on the \emph{Targeted Advertising} scenario, where entertainment or infotainment content is enriched with advertising objects and hyperlinks. For both scenarios, first concepts for potential pilots have been developed during joint workshops with the broadcasters production department. The most promising concepts have been refined subsequently and transformed into storyboards, which formed the base for the actual pilot production. Two pilots have been produced by RBB: \emph{T\"ater Opfer Polizei} and \emph{Fontane360}. The first one is a CSI-like crime scene investigation with gamification elements, whereas the second one is more calm and follows Fontane through one of his walks through the Mark Brandenburg. RTI produced also two pilots named \emph{TechnoGym} and \emph{Universita Cattolica of Milan}. The former explains the proper use of gym equipment, whereas the latter has the aim of helping new students in choosing the right course. The pilots were used subsequently to asses the whole immersive experience as well as its major components (like the players, 3D mentor, personalisation, automatic camera path) by test consumers. The gathered feedback from the test consumers was generally positive.

\section{Conclusion}

The work done so far in the Hyper360 project on tools for capturing, production, enhancement, delivery and consumption of enriched 360$^\circ$ video content has been presented. Furthermore, the first pilots which have produced with these tools have been described.

\section*{Acknowledgment}

This work has received funding from the European Union's Horizon 2020 research and innovation programme, grant n$^\circ$ 761934, Hyper360 (``Enriching 360 media with 3D storytelling and personalisation elements''). Thanks to Rundfunk Berlin-Brandenburg and Mediaset for providing the 360\textbf{$^\circ$}~video content.

\balance

\bibliographystyle{IEEEbib}
\bibliography{icme2020template}

\end{document}